\documentclass[prl,twocolumn,showpacs,preprintnumbers,amsmath,amssymb]{revtex4}

\usepackage{graphicx}
\usepackage{dcolumn}
\usepackage{bm}

\begin{document}

\preprint{PRL Preprint}

\title{Fermionic light in common optical media}
\author{David Novoa, Humberto Michinel and Daniele Tommasini}
\affiliation{Departamento de F\'{\i}sica Aplicada, Facultade de Ciencias de Ourense,\\
Universidade de Vigo, As Lagoas s/n, Ourense, ES-32004 Spain.}

\begin{abstract}
  Recent experiments have proven that the response to short laser pulses of
  common optical media, such as air or Oxygen, can be described by focusing
  Kerr and higher order nonlinearities of alternating signs. Such media
  support the propagation of steady solitary waves. We argue by both numerical
  and analytical computations that the low power fundamental bright solitons
  satisfy an equation of state which is similar to that of a degenerate gas of
  fermions at zero temperature. Considering in particular the propagation in
  both $O_2$ and air, we also find that the high power solutions behave like
  droplets of ordinary liquids. We then show how a grid of the fermionic light
  bubbles can be generated and forced to merge in a liquid droplet. This leads
  us to propose a set of experiments aimed at the production of both the
  fermionic and liquid phases of light, and at the demonstration of the
  transition from the former to the latter.
\end{abstract}


\maketitle

In suitable optical media, light has been argued to acquire material
properties. A long dated example is the equivalence of the paraxial
propagation of a laser pulse in a Kerr medium with the time evolution of a
superfluid Bose-Einstein condensate, due to the identity of the nonlinear
Schr\"odinger equation with the Gross-Pitaevskii equation{\cite{GP-NLSE}}.
More recently, optical induction has been used to create photonic
crystals{\cite{Photonic-crystals}}; a photonic system has been designed that
may undergo a Mott insulator to superfluid quantum phase
transition{\cite{Solid-light}}; 
soliton solutions for light propagation in Cubic-Quintic (CQ)
nonlinear media have been shown to 
behave like ordinary liquids{\cite{michinel02,novoa09}}.

On the other hand, recent experimental and theoretical works
have proven that the response to ultrashort laser pulses of common optical
media, such as air or Oxygen, can be described by focusing Kerr\cite{roso} and higher
order nonlinearities of alternating signs\cite{Oxygen}, which have also been 
argued to provide the main mechanism in
filament stabilization, instead of the plasma defocusing\cite{Oxygen}.

In this letter, we demonstrate by analytical and numerical computations 
that such media can support the propagation of steady solitary waves that appear in 
two clearly different phases. The low power solitons are governed by the same equation of
state than a degenerate gas of fermions. We will call such a system
\emph{``Fermionic Light''}. On the other hand, the high power localized
states satisfy the Young-Laplace equation that governs the formation
of droplets in ordinary liquids, similarly to the result that was recently
obtained for the CQ model{\cite{novoa09}}. 
We also show how to generate a grid of fermionic light bubbles and make it
turn into a liquid light droplet.

We will consider the (paraxial) propagation along the $z$-direction of a linearly
polarized laser beam, so that the complex electric field component $\Psi
(x, y, z)$ satisfies the nonlinear Schr\"odinger equation
\begin{equation}
  \label{NLSE} i \frac{\partial \Psi}{\partial z} + \frac{1}{2 k_0 n_0}
  \nabla_{\perp}^2 \Psi + k_0 \Delta n \Psi = 0,
\end{equation}
where $n_0$ is the linear refractive index of the medium, $\nabla_{\perp}^2 =
\partial^2 / \partial x^2 + \partial^2 / \partial y^2$ is the transverse
Laplace operator, and $k_0 = 2 \pi / \lambda_0$ is the mean wavenumber in
vacuum, where $\lambda_0$ is the central wavelength of the laser source which
will be fixed to $\lambda_0 = 800 nm$ throughout this paper as in the
experiment of Ref.{\cite{Oxygen}}. For the optical media that have been
studied in Ref.{\cite{Oxygen}}, the nonlinear correction $\Delta n$ to the
refractive index can be expanded as
\begin{equation}
  \label{index} \Delta n \simeq n_2 | \Psi |^2 + n_4 | \Psi |^4 + n_6 | \Psi
  |^6 + n_8 | \Psi |^8,
\end{equation}
with alternating sign coefficients $n_2, n_6 > 0$ and $n_4, n_8 < 0$, that
contribute to focusing and defocusing respectively. Taking into account that the values 
of the second-order dispersion and multiphoton-absorption coefficients for the air are 
$k''=0.2$ $fs^2/cm$ and $\beta=1.27\times10^{-126}$ $cm^{17}/W^9$\cite{nl_processes}, respectively, 
we have checked that both effects do not lead to significant corrections in our results. 
To be concrete, we will perform
most of the numerical calculations in the case of $O_2$ as the propagation
medium, taking the mean values obtained in the experiment\cite{Oxygen}, 
$n_2 = 1.6 \times 10^{- 19} {cm}^2 / W$,
$n_4 =- 5.2 \times 10^{- 33} cm^4 / W^2$, $n_6 = 4.8 \times
10^{- 46} cm^6 / W^3$, $n_8 =- 2.1 \times 10^{- 59}
cm^8 / W^4$. For comparison, however, we will also mention
the results that we obtain for air, using the
corresponding values for the coefficients $n_{2 q}$ that are also
given in Ref.{\cite{Oxygen}}.

\begin{figure}[htbp]
{\centering \resizebox*{1\columnwidth}{!}{\includegraphics{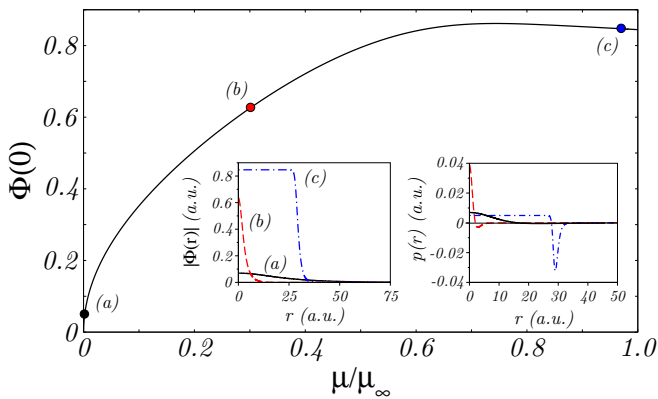}} \par}
\caption{[Color online] Plot of the central amplitude $\Phi(0)$ vs. $\mu/\mu_\infty$ 
for the family of fundamental eigenstates of Eq.\eqref{NLSE}. Note that $\Phi(0)$ 
has a maximum for $\mu/\mu_\infty\approx 0.7$, due to the defocusing nonlinearities. 
Left inset: radial 
profiles of three eigenfunctions of Eq.\eqref{NLSE} with $\mu/\mu_\infty=0.004$ (black solid), 
$\mu/\mu_\infty=0.3$ (red dashed) and $\mu/\mu_\infty=0.97$ (blue dashed-dotted) respectively. 
Right inset: radial pressure profiles corresponding to these solutions. 
The low-power profile is magnified by a factor $10^3$ for 
clarity. Labeled points on the main curve refer 
to the eigenstates displayed within the insets.}
\label{fig1}
\end{figure}


We will search for finite localized solutions of Eq. \eqref{NLSE} of the form $\Psi (x,
y, z) = \Phi (r) e^{- i \mu z}$, where $r = \sqrt{x^2 + y^2}$ and $\mu$ is the
propagation constant. Fig. 1 shows the result of our numerical computation for
$\Phi (0)$ in the existence domain of such solitons in Oxygen ($\mu_{\infty} < \mu
< 0$), where $\mu_{\infty}=$ $- 0.096$ $cm^{-1}$, and for the radial profiles $\Phi (r)$ of three of them (see
left inset in Fig.\ref{fig1}), corresponding to the low power (black solid
line), moderate power (red dashed line) and high power regimes (blue
dashed-dotted line) respectively. We express the transverse spatial variables $x,y$ in terms of the 
adimensional coordinates $\xi,\chi$, measured in units of $(n_4/n_0)^{1/2}(k_0n_2)^{-1}$.  
The amplitude $|\Phi (0)|$ is measured in units of $(n_2/n_4)^{1/2}$. 

Like in the CQ case, $\mu$ can be identified with the chemical potential of an
equivalent thermodynamical two-dimensional system{\cite{novoa09}} of $N = \int
\rho dxdy \equiv \int | \Phi |^2 dxdy$ particles, described by the Landau's
grand potential{\cite{Landau}} $\Omega = - \int pdxdy$, where the pressure
field $p$ is now
\begin{equation}
  p = - \frac{1}{2 k_0 n_0} | \nabla_{\perp} \Phi |^2 + \mu | \Phi |^2 + k_0
  \sum_{q = 1}^4 n_{2 q} \frac{\Phi^{2 (q + 1)}}{q + 1} \label{pressure}
\end{equation}
For our optical system, $\mu$, $N$, $\Omega$ and $p$ correspond to the propagation constant, the power, 
the Lagrangian leading to Eq. \eqref{NLSE}, and the Lagrangian density, respectively.

The right inset of Fig.\ref{fig1} shows the pressure distributions $p(r)$, measured in 
units of $k_0n_2^3n_4^{-2}$, corresponding to the stationary states discussed above. 
Notice that all the radial distributions display both positive and negative pressure regions,
being this a necessary condition for the existence of solitary waves. 
For the low power soliton (black solid line), corresponding to small values of $| \mu
|$, we can obtain an analytical expression by using the variational method
with the ansatz $\Phi (A, r) = A \exp (- r^2 / R^2)$. The values of $A (\mu)$
and $R (\mu)$ that minimize $\Omega$ for a given value of $\mu$ can then be
used to compute the pressure distribution. Taking the first non-vanishing
order in $| \mu |$ and inverting the dependence $R (\mu)$, we find a simple
analytical approximation for the central pressure $p_c$ as a function either
of the radius $R = \sqrt{2 \langle r^2 \rangle}$ of the soliton, or of the
central density $\rho_c = | \Phi (0) |^2$ (corresponding to the beam intensity),
\begin{equation}
p_c = \frac{a}{k_0^3 n_0^2 n_2} \frac{1}{R^4} = bk_0 n_2 \rho_c^2,
\label{pressure_fermilight}
\end{equation}
with $a = 4$ and $b = 1/4$. This relation is similar to the equation of
state of a degenerate gas of fermions of mass $m$ at zero temperature. In
fact, if we apply the general definition of the Fermi momentum{\cite{Landau}}
$P_F$ to a two dimensional system, we obtain $P_F = \hbar \sqrt{2 \pi \rho}$,
with $\rho$ the density of the Fermi gas. As a consequence, the pressure,
defined as the average force on a unit orthogonal line in the gas, can be
obtained from the average kinetic energy as follows
\begin{equation}
  p = \rho \langle E_{kin} \rangle = \frac{\rho}{2 m}
  \frac{\int_0^{P_F} P^2 PdP}{\int_0^{P_F} PdP} = \frac{\pi \hbar^2}{2 m}
  \rho^2
\end{equation}
which shows the same dependence with $\rho^2$ as Eq.  \eqref{pressure_fermilight}. 
This proves the formal
analogy of our low power solitons with a degenerate Fermi gas in the central
region around $r = 0$, which is arbitrarily large in the limit $\mu \to 0$
(corresponding to large radius $R \to \infty$). For these reasons, we will
call \emph{'Fermionic'} the phase where the pressure is proportional to
$\rho^2$.

\begin{figure}[htbp]
{\centering \resizebox*{1\columnwidth}{!}{\includegraphics{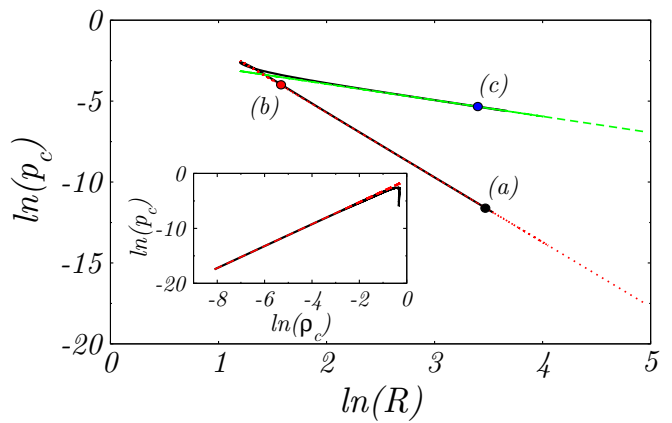}} \par}
\caption{[Color online]
Plot of the logarithms of $p_c$ vs. $R$ for the fundamental solitons (black line). 
The fermionic behavior of Eq.\eqref{pressure_fermilight} (red dotted) 
and the liquid YL equation (green dashed) are compared with the numerics. 
The labeled points correspond to the same
eigenstates displayed in Fig.\ref{fig1}. For each value of $R$, two different outcomes 
are possible depending on the power, corresponding to two different phases of light. 
Inset: plot of $ln(p_c)$ vs. $ln(\rho_c)$ (black solid) overlapped with Eq.\eqref{pressure_fermilight}
(red dashed) for comparison.}
\label{fig2}
\end{figure}

Note that in the limit $\mu \to 0$ our variational computation gives a
constant $N = \frac{2 \pi}{k_0^2 n_0 n_2}$, which is consistent with the known
result for the power flow leading to the collapse threshold in a Kerr
medium{\cite{Marburger75}}. The magnitude of this power $N$ lies in the range of few $GW$ 
in both $O_2$ and air, and can be interpreted as the threshold for the existence of the 
Fermionic Light solitons.

Fig.\ref{fig2} shows the numerical computation of $p_c$ as a function either
of $R$ (black solid line) or $\rho_c$ (inset: black solid line) for all the
nodeless solitary states of the model. The lower branch, corresponding to the
low power solitons, is in excellent agreement with the dependence described by
Eq.\eqref{pressure_fermilight}, as it can be inferred from the fitting (red dotted) straight line
with slope $s_{low} = - 4$. Furthermore, in the inset of Fig.\ref{fig2} we
also show the quantitative agreement between theory and numerics by comparing
$p_c$ with $\rho_c$ instead of $R$. In this case, the slope of the straight
line is $s_{inset} = 2$ thus demonstrating the quadratic dependence on
$\rho_c$ given by Eq.\eqref{pressure_fermilight}. However, the correct numerical values of the
constants are $a = 2.5$ and $b = 0.29$, in reasonable agreement with the
result of the variational method given above. We have checked numerically that
the asymptotic behavior represented by the red lines in Fig.\ref{fig2} is
practically independent on the higher order nonlinearities $n_{2 q}$ $(q >
1)$, as predicted by the theory. In particular, these results can be directly
applied to both Kerr and CQ models.

On the other hand, the high power localized solutions exhibit top-flat
profiles with an inhomogeneous negative pressure profile on the border,
similar to those of the solitons appearing in the CQ model\cite{novoa09}. As
a consequence, the gradient term in Eq.\eqref{NLSE} can be neglected close to the origin,
and we get $\mu = - k_0 \Delta n (0) = - k_0 \sum_{q = 1}^4 n_{2 q} \Phi
(0)^{2 q}$. By generalizing the argument of Ref.\cite{novoa09} including the higher order nonlinearities, 
we obtain that these states obey
the celebrated Young-Laplace (YL) equation\cite{surften_review}, $p_c = 2
\sigma / R$, describing the behavior of usual liquid droplets. The value of
the surface tension is
\begin{equation}
  \sigma = \frac{n_2}{\sqrt{2 n_0 n_4}} \int^{\Phi_{\infty}}_0 \left( -
  \mu_{\infty} \Phi^2 - k_0 \sum_{q = 1}^4 n_{2 q} \frac{\Phi^{2 (q + 1)}}{q +
  1} \right)^{1 / 2} d \Phi,
\end{equation}
being $\mu_{\infty}$ and $\Phi_{\infty}$ the asymptotic values corresponding
to the $R \to \infty$ droplet, that can be computed by solving the equation
$p_c = 0$ (neglecting the Laplacian term). For the propagation in Oxygen,
we have obtained the following
variational estimations: $\mu_{\infty} = - 0.247 (k_0 n_2^2
n_4^{- 1}) = - 0.096$ $cm^{- 1}$, $\Phi_{\infty}^2 = 0.712 (n_2 n_4^{- 1})
= 2.19 \times 10^{13}$ $W / cm^2$, $\sigma = 0.0715 (n_2^2 n_4^{- 3 / 2}
n_0^{- 1 / 2}) = 4.88 \times 10^9$ $W / cm^2$.
\begin{figure}[htbp]
{\centering \resizebox*{1\columnwidth}{!}{\includegraphics{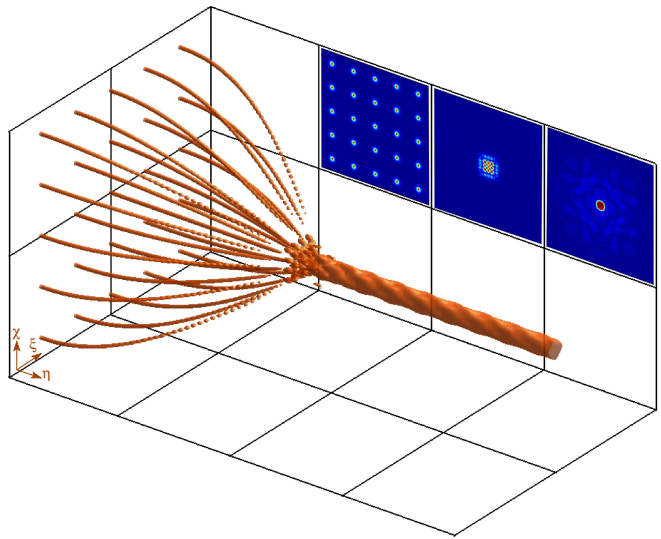}} \par}
\caption{[Color online]
Isosurface intensity plot of the dynamical phase transition from a square grid of fermionic light solitons to 
an individual liquid light soliton in $O_2$. All solitons in the grid are eigenstates with $\mu/\mu_\infty=0.3$ displayed 
in Fig.\ref{fig1}. They are forced to merge in the center of the computational window by means of an external 
harmonic potential. We can see three pseudocolor plots displaying the initial condition (left, $\eta=0$), 
the collapse of the light bubbles (middle, $\eta=200$) and the final state (right, $\eta=500$). In the latter we 
observe a flat-top soliton with radius $R\approx10$ and 
$\rho_c\approx0.76$ arising after the massive coalescence of the fermionic 
solitons. The width of the square spatial domain displayed is $w_{\xi,\chi}=200$.}
\label{fig3}
\end{figure}
These results, obtained assuming a top-flat function, are in excellent
agreement with the computation given in Fig.\ref{fig1}. In
Fig.\ref{fig2}, we show that our numerical solution in the case of $O_2$
satisfy the YL equation (green dashed line) with very good accuracy for a wide
range of values of $\mu$.
On the other hand, for the propagation in air, the liquid light phase would
correspond to a higher intensity, $[\Phi_{\infty}^2]_a = 2.98 \times 10^{13} W
/ cm^2$, with $[\mu_{\infty}]_a = - 0.116$ $cm^{- 1}$ and $[\sigma]_a = 7.16
\times 10^9 W / cm^2$, and the YL equation would still be valid.

As we have seen above, the propagation of self-guided light beams in media like
Oxygen can occur in two clearly separated phases, satisfying two different
equations of state. In fact, we have checked that qualitatively similar results 
can also be obtained for the CQ case, and occur whenever the nonlinear refractive index 
displays a single well-defined maximum as a function of the intensity.
It would then be interesting to demonstrate the
possibility of a transition between the fermionic bubbles and the liquid
droplets of light. A suggestive analogy is that of the collapse of a star,
that occurs when the gravitational interaction overcomes the Fermi pressure of
the electrons. We can obtain a qualitatively similar result in the case of
light propagation in Oxygen, by compressing the fermionic bubbles using a
harmonic potential, leading to the generation of a liquid light droplet.
Fig. 3 shows the result of our simulation in $O_2$. The initial state (see left
snapshot in Fig. \ref{fig3}) consists of a regular grid of fermionic light
bubbles with $\mu / \mu_{\infty} = 0.3$ (see their radial profile in Fig.
\ref{fig1}), with a separation between nearest neighbors $\Delta_{\xi,\chi}=40$. We include an external harmonic potential 
$V (\xi, \chi) = \frac{K}{2} (\xi^2 + \chi^2)$ with $K = 5 \times 10^{- 5}$, 
in Eq.\eqref{NLSE} in order to induce a net force acting on the grid with the aim of 
making all the fermionic solitons to collide in the center of the computational 
window $(\xi, \chi) = 0$. 

This parabolic potential can be obtained by inducing in the medium a "gas lens"\cite{gas_lens}, 
which can be constructed with an electrically heated pipe through which passes a laminar flow of gas. By controlling the differential heating at the boundaries and the velocity of the flow, a parabolic refractive index gradient can be obtained 
as documented in Ref.\cite{gas_lens}. On the other hand, we have also checked numerically that the same qualitative result
can  be obtained by using a glass lens, provided that its focal length $f$ is
at least ten times greater than the Rayleigh length of the input beams, which in the case depicted in Fig. 3 corresponds to 
$f$ $=$ $5$ $m$.

\begin{figure}[htbp]
{\centering \resizebox*{1\columnwidth}{!}{\includegraphics{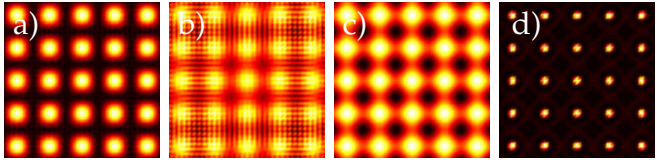}} \par}
\caption{[Color online] Proposal for the generation of the initial condition of 
Fig.\ref{fig3} in experimental setups. The (a) pseudocolor plot shows the bidimensional 
cosine squared-like phase distribution to be imprinted onto the ingoing beam. We assume 
that the beam is much wider than the filaments, 
and restrict our simulation to a square domain $\xi,\chi\in[-75,75]$ where the 
wavefront is assumed to be homogeneous. In snapshots (b)-(c) ($\eta=100,200$) we observe how the 
beam profile starts to destabilize, leading to the formation of a grid of spatial solitons at $\eta=300$ 
(see snapshot (d)).}
\label{fig4}
\end{figure}

For simplicity, we have prepared the initial state with exact solutions of Eq.\eqref{NLSE} to reduce the
excitation of linear radiative modes. 
We have checked that such an initial guess can be generated
experimentally by means of the modulational instability of a low power probe
pulse. In fact, Fig.\ref{fig4} shows the results of a numerical simulation where we
have added an initial cosine squared-type phase distribution (see left snapshot in
Fig.\ref{fig4}) to a homogeneous plane wave in order to excite a regular grid
of solitons, 
as in Refs. \cite{filament_control}.
As shown in snapshots b) to d), 
this allows us to control the spatial location of the optical
filaments generated during the beam breakup due to modulational instability. 
The outgoing grid of spatial solitons is depicted
in snapshot d). Note its similarity with the initial state of 
Fig.\ref{fig3}.

Let us come back to the results of Fig.\ref{fig3}. The massive
coalescence of all the fermionic bubbles occurs after a finite propagation
distance $\eta = 200$ (see middle snapshot in Fig.\ref{fig3}), 
measured in units of $n_4k_0^{-1}n_2^{-2}$. As a result, a
unique filament structure with large radius arises, as it can be appreciated
in Fig.\ref{fig3}. We have checked that this soliton is a flat-top eigenstate
with radial perturbations coming from the transition process. In fact, we have
estimated its logarithmic radius to be around $2.3$ and its peak density
$\rho_c \approx 0.76$  (measured in units of $n_2/n_4$, thus corresponding with an 
intensity $2.34\times10^{13}$ $W/cm^2$), which is clearly on the liquid light branch displayed
in Fig.\ref{fig2}. For Oxygen\cite{Oxygen}, we have considered a
propagation distance of about $13$ m. Although such a distance seems to be
huge for a real experiment, we have verified that by stretching the external
potential this distance can be realistically reduced by almost one order of
magnitude. Finally, we note that the use of pressurized $O_2$ or air may also help to
enhance the nonlinear optical response. For all these reasons, we conclude
that the demonstration of the existence of both Fermionic and Liquid light and
the phase transition between them would be an affordable challenge in real
experiments.

In conclusion, we have proven that common media ($O_2$, air) can support the
propagation of solitary waves that appear in two clearly different phases with
unequal physical properties, namely the low power \emph{``Fermionic''}
light, satisfying an equation of state similar to that of a degenerate gas of
fermions, and the high power \emph{``Liquid''} light, obeying the YL
equation. We have then shown how a grid of the fermionic light bubbles can be
generated and forced to merge in a liquid droplet. 
We think that the possible experimental validation of our proposal 
could also provide an independent way to corroborate 
the deep change in the understanding of the
filamentation process in gases that was proposed in Ref.\cite{Oxygen}.
Furthermore, these results in air pave the way for the improvement of recent experiments on 
laser-induced water condensation\cite{rain}, 
built on top of these new robust light distributions.

This work was supported by MICINN, Spain (project FIS2008-01001). D.N. acknowledges support 
from Conseller\'{i}a de Econom\'{i}a e Industria-Xunta de Galicia through 
the ``Maria Barbeito'' program.


\begin{thebibliography}{99}

\bibitem{GP-NLSE} 
P. G. Kevrekidis, D. J. Frantzeskakis and R. Carretero-Gonzalez, Eds., \emph{Emergent Nonlinear Phenomena in 
Bose-Einstein Condensates} (Springer-Verlag, 2008).

\bibitem{Photonic-crystals} 
T. Pertsch, P. Dannberg, W. Elflein, A. Brauer, F. Lederer,
\prl {\bf 83}, 4752 (1999); 
R. Morandotti, U. Peschel, J. S. Aitchison, H. S. Eisenberg, Y. Silberberg,
\prl {\bf 83}, 4756 (1999). 
H. Martin, E. D. Eugenieva, Z. Chen, D. N. Christodoulides,
\prl {\bf 92}, 123902 (2004);
B. Freedman {\it et al}, Nature {\bf 440}, 1166 (2006);
T. Schwartz {\it et al}, Nature {\bf 446}, 52 (2007).

\bibitem{Solid-light} 
A. D. Greentree, C. Tahan, J. H. Cole and L. C. L. Hollenberg,
Nature Physics {\bf 2}, 856 (2006).

\bibitem{michinel02} H. Michinel, J. Campo-Taboas, R. Garcia-Fernandez, J. R. Salgueiro, M. L. Quiroga-Teixeiro,
\pre {\bf 65}, 066604 (2002).

\bibitem{novoa09} D. Novoa, H. Michinel and D. Tommasini,
\prl {\bf 103}, 023903 (2009).

\bibitem{roso} C. Ruiz {\it et al},
\prl {\bf 95}, 053905 (2005).

\bibitem{Oxygen} 
V. Loriot, E. Hertz, O. Faucher, and B. Lavorel, Opt.
Express {\bf 17}, 13429 (2009); Opt. Express {\bf 18}, 3011(E)
(2010);
P. Bejot {\it et al.}, 
\prl {\bf 104}, 103903 (2010).

\bibitem{nl_processes} S. Tzortzakis {\it et al},
\prl {\bf 86}, 5470 (2001).

\bibitem{Marburger75} J. H. Marburger,
Prog. Quant. Electr. {\bf 4}, 35 (1975).

\bibitem{Landau} L. D. Landau and E. M. Lifshitz,
\emph{Statistical Physics}, (Pergamon, Oxford, 1984).

\bibitem{surften_review} J. Eggers,
\rmp {\bf 69}, 3 (1997).

\bibitem{gas_lens} Notcutt, M. {\it et al}, 
Opt. Laser Technol., {\bf 20}, 243 (1988);
Michaelis, M. M. {\it et al},
Nature {\bf 353}, 547 (1991).

\bibitem{filament_control} H. Schroeder, J. Liu, and S. L. Chin,
Opt. Express {\bf 12}, 4768 (2004);
M.R. Fetterman {\it et al.}, Opt. Express {\bf 3}, 366 (1998).

\bibitem{rain} P. Rohwetter {\it et al}.,
Nature Photonics, {\bf 4}, 451 (2010). 

\end{thebibliography}
\end{document}